\input epsf
\def\beq{\begin{equation}}
\def\eeq{\end{equation}}
\def\bey{\begin{eqnarray}}
\def\eey{\end{eqnarray}}
\def\kms{\mbox{\rm \,km\,s}^{-1}}
\documentstyle[11pt,paspconf]{article}

\begin{document}

\title{Studying Evolution of the Galactic Potential and Halo Streamers with 
Future Astrometric Satellites}


\author{
HongSheng Zhao\altaffilmark{1}, 
Kathryn V. Johnston\altaffilmark{2}, 
David N. Spergel\altaffilmark{3} 
\& Lars Hernquist\altaffilmark{4}}

\altaffiltext{1}{Sterrewacht Leiden, Niels Bohrweg 2, 2333 CA,
Leiden, The Netherlands}
\altaffiltext{2}{Institute for Advanced Study, Princeton, NJ 08450}
\altaffiltext{3}{Princeton University Observatory, Princeton University,
Princeton, NJ 08540}
\altaffiltext{4}{Department of Astronomy, Harvard University,
Cambridge, MA 02138.}

\begin{abstract}
Future astrometric satellites, such as SIM (NASA's Space
Interferometric Mission) and GAIA (ESA's Global Astrometric
Interferometer for Astrophysics), hold the promise of mapping out the
detailed phase space structure of the Galactic halo by providing
unprecedented annual proper motion and parallax of $1-10\mu$as
astrometric accuracy.  Here we show that proper motions of hundred or
so giant branch stars in a tidal debris torn from a small satellite (a
$10^{5-7}L_\odot$ Galactic dwarf galaxy or globular cluster) in the
halo is sensitive to the current Galactic potential and its past
evolution.  We follow the evolution of a cold (velocity dispersion of
10 km/s) stream on a nearby (between 8-50 kpc) polar orbit in a
variety of histories of the potential of the Galaxy, and observe the
bright ($V<18$mag) members of the debris tail with GAIA accuracy.  We
simulate effects due to the growing or flipping of the Galactic disk
over the past 4 Gyrs or the perturbation from a massive accreted lump
such as the progenitor of the Magellanic Clouds.  We study various
factors influencing our ability to identify streams, including
contamination from field stars, accuracy of radial velocity and
distance data and evolution and non-axial symmetry of the potential.
Our simulations suggest that nearby, cold streams could be detected
with GAIA if these cousins of the Sagittarius stream exist.  Results
of Johnston, Zhao, Spergel \& Hernquist (1999) and Helmi, Zhao \& de
Zeeuw (1999) for static Galactic potentials are likely to be largely
generalizable to moderately time-evolving potentials.  SIM and GAIA
measurements of debris stars might be used to probe both Galactic
structure and Galactic history.  
\end{abstract}

\section{Introduction}

Tidal streams in the Galactic halo are a natural prediction of
hierarchical galaxy formation, where the Galaxy builds up its mass by
accreting smaller satellite galaxies.  They are often traced by
luminous horizontal and giant branch stars outside the tidal radius of
a satellite (such as the Sagittarius dwarf galaxy) or a globular
cluster (cf. Grillmair et al. 1998, Irwin \& Hatzidimitriou 1995) as a
result of tidal stripping, shocking or evaporation.  That
extra-tidal material (stars or gas clouds) traces the orbit of the
satellite or globular cluster has long been known to be a powerful
probe of the potential of the Galaxy in the halo, and has been
exploited extensively particularly in the case of the Magellanic
Clouds and Magellanic Stream (Murai \& Fujimoto 1980, Putman et
al. 1999, this volume).

Among the future astrometric missions, SIM is a pointed observatory
with the main science goal of finding nearby planets, while GAIA
(under study) will continuously monitor the relative positions of about
$10^9$ stars above $\sim 20$ mag. limit over the full sky over a
period of $4-5$ years with additional information of radial velocity
to better than $3-10$km/s accuracy (depending on spectral type) for
stars brighter than $\sim 16$ mag. (cf. the compiled instrument specs
in Table 1, and the original documents by Gilmore et al. 1998, Unwin
et al. 1998). GAIA represents an improvement over the Hipparcos
mission by a factor of about a thousand in precision, more than a
million in the probing volume, and a factor of ten thousand in number
of objects.

Helmi, Zhao \& de Zeeuw (1999, this volume) show that streams can be
identified by as peaks in the distribution in the angular momentum
space, measurable with GAIA.  Once identified, we can fit a stream
with an orbit or more accurately a simulated stream in a given
potential.  Johnston, Zhao, Spergel \& Hernquist (1999) show that a
few percent precision in the rotation curve, flattening and
triaxiality of the halo is reachable by mapping out the proper motions
(with SIM accuracy) and radial velocities along a tidal stream $\> 20$
kpc from the Sun.  In particular, they show that the fairly large error
in distance measurements to outer halo stars presents no serious
problem since one can predict distances theoretically using the known
narrow distribution of the angular momentum or energy along the tails
associated with a particular Galactic satellite.  We expect these
results should largely hold for streams detectable by GAIA.  These
numerical simulations are very encouraging since they show that it is
plausible to a learn great deal about the Galactic potential with even
a small sample of stream stars from GAIA.  Some unaddressed issues
include whether stream members will still be identifiable in angular
momentum in potentials without axial symmetry, and the robustness of
both methods if the Galactic potential evolves in time.

Here we illustrate how the properties of tidal streams evolve in 
time-dependent potentials and discuss whether members of
such streams might still be identified
using the 6D information from GAIA.
We study the effect of contamination from field stars, evolution and non-axial
symmetry of the potential, and the lack of very accurate radial
velocity and parallax on our ability to detect
streams.  We concentrate on satellites which fall in and are disrupted
recently (about 4 Gyrs ago, well after the violent relaxation phase)
and maintain a cold spaghetti-like structure.  We also put the
satellites on relatively tight orbits but which lie
outside the solar circle
(pericenter of about 8 kpc and apocenter of about 40 kpc) such that
the bright member stars in the stream are still within the reach of
detectability of GAIA.  Such streams typically go around the Galaxy
less than 5 times since disruption, and are typically far from
fully phase-mixed.

\begin{table}
\caption{Preliminary specifications of SIM and GAIA, as adopted in this paper}
\begin{center}
\begin{tabular}{lll}
   &SIM &GAIA\\
program   & NASA origin & ESA (study)\\
launch + duration & 2006 + 5 & 2009 + 5 \\
orbit & Sun-synchronous ($100$min) & L2 \\
resolution ($mas$)      & $\sim 10$ & $\sim 100$ \\
parallax ($\mu as$)      & 4 & 4-200 for $V \le 10-20$ mag.\\
proper motion ($\mu as$yr$^{-1}$) & 1 & 4-200 for $V \le 10-20$ mag. \\
radial velocity (${\rm km s}^{-1}$) &  & 3-10 down to $V \sim 16$ mag.\\
wavelength($\mu m$) &0.4-1   & 0.4-1 \\
objects (mag. limit) & $10^4$ ($V\sim 20$) & $10^9$ ($V\sim 20$) \\
coverage & pointed, narrow/wide field  & full sky scan
\end{tabular}
\end{center}
\end{table}

\section{Strategy for GAIA: stars in a stream vs. field stars}

We propose to select bright horizontal/giant branch (HB/GB) stars as
tracers of tidal debris of a halo satellite (which we take to be either a
dwarf galaxy or a globular cluster).  A satellite with a
typical luminosity $L=10^{5-7}L_\odot$ has numerous HB and GB stars
with $M_V \le 0.75$mag, which are easily observable at 20 kpc from the
Galactic center ($m_V \approx 18$mag or brighter) with GAIA.  While
the depth of the debris is typically difficult to resolve with GAIA
parallax, GAIA proper motion (cf. Table 1) is good enough to resolve
the internal dispersion of the debris, which is of the order
$\sigma/20{\rm kpc} \approx 100\mu$as/yr, where $\sigma \sim 10\kms$ is the
typical velocity dispersion of a satellite.

A satellite with luminosity $10^{5-7}L_\odot$ typically has
\beq
N_{obs}={fL \over 500L_\odot} \ge 100
\eeq
numbers of stars in its tidal tail, where we assume there is about one 
HB or GB star per $500L_\odot$ and between $f=0.5\%$ to $50\%$ of the stars 
in the original satellite are liberated.

In comparison, the density of field HB or GB stars in the halo
which happen to be in the same solid angle, and have the same proper motion
and radial velocity
\beq
N_{field}=\Sigma \Omega \left( {\sigma \over 100\kms} \right)^3 \sim 1
\eeq  
where $100\kms$ is the dispersion of field stars, $\Sigma \sim (1-10)$
is the typical number of field giants per square degree and $\Omega$
is the solid angle of the tidal debris in degrees, which for the
Sagittarius (Sgr) tidal stream is about $5^o\times (20^o-60^o)$.  In
other words while there might be 25 stars in a solid angle of
$5^o\times 5^o$ in a stream, only $25\%$ of the chance that there is
one field star in the same piece of sky with proper motions and radial
velocities indistinguishable from the stream stars.  In fact Sgr was
discovered on the basis of radial velocity and photometric parallax
against a dense foreground of bulge stars (Ibata, Gilmore, \& Irwin
1994).  So as far as identifying stars in a cold stream with GAIA is
concerned we conclude that contamination from field halo stars is
likely not a serious problem.

It is worth commenting on the advantages of stars in a stream, as
compared to stars in the field, in constraining the Galactic
potential.  Stars in a stream trace a narrow bunch of orbits in the
vicinity of that of the center of the mass, and are correlated in orbital
phase: they can all be traced back to a small volume (e.g., near
pericenters of the satellite orbit) where they were once bound to the
satellite.  Hence we expect a tight constraint on parameters of the
Galactic potential and the initial condition of the center of the mass
of the satellite (about a dozen parameters in total) by fitting the
individual proper motions of one hundred or more stars along a stream
since the fitting problem is over-constrained.  In contrast, field
stars are random samples of the distribution function (DF) of the
halo, and the large number of degrees of freedom in choosing the 
6-dimensional
DF makes the problem often under-constrained: one generally cannot
make the assumption that the halo field stars are in steady state as
an ensemble because it typically takes much longer than a Hubble time
to phase-mix completely at radii of 30 kpc or more.

How well can we determine the Galactic potential with a stream?
Assume each debris star has an intrinsic energy spread $\Delta
E \sim V_{cir}^2 a/R$ relative to the orbit of the center of the mass, 
where $a=(0.1-1)$ kpc is the size of the parent satellite
at the time of disruption, also the thickness of a cross-section of the stream,
and $R=(8-40)$ kpc is the radius of the pericenter,
then the accuracy of the potential from fitting $N_{obs}\sim (100-1000)$ stars 
is
\beq 
\epsilon \sim {1 \over \sqrt{N_{obs}} } {\Delta E \over V_{cir}^2} \sim 
(0.1-1)\% 
\eeq 
The accuracy is not very sensitive to the luminosity of the disrupted 
satellite since 
both $N_{obs}$ and $a$ scale with the luminosity of the satellite.
While this agrees fairly well with the study of parameterized 
static potentials 
(Johnston et al. 1999), it remains to be tested for more realistic models
with evolution.

\section{Science for GAIA: evolution of the Galactic potential}

To study the effect of the evolution and flattening of the potential
on a stream, we follow the disruption of satellites in a simple,
flattened, singular isothermal potential $\Phi(r,\theta,t)$ which is
time-dependent but maintains a rigorously flat rotation curve at all
radii, where $(r,\theta)$ are the spherical coordinates describing the
radius and the angle from the North Galactic Pole, $t$ is defined such
that $t=0$ would be the present epoch.  This is perhaps a sensible
choice since we do not expect a major merger to have occurred within
the past 4 Gyrs.  We
study three types of moderately evolving potential with identical
flattening and rotation curves at the present epoch.

First we consider a Galactic potential 
\beq
\Phi_G(r,\theta,t) = V_0^2 \left[A_s \log r + 
{\epsilon \over 2} \cos 2\theta \right],
\eeq
where 
\beq
A_s(t)=1-\epsilon_0+\epsilon(t),
~\epsilon(t)=\epsilon_0 \cos{2\pi t \over T_G}.
\eeq
This model simulates the effect of the Galaxy becoming
more massive and flattened in potential as it grows a disk.
The time-evolution is such that 
the Galactic potential grows from prolate at time $t=-T_G/2$
to spherical at time $t=-T_G/4$, and then to oblate at $t=0$.
A more general prescription of the temporal variation 
might include a full set of Fourier terms.  
We adopt parameters 
\beq
V_0=200\kms, ~~~ \epsilon_0=0.1, 
\eeq
such that the present-day rotation curve amplitude is $200\kms$, and
equal potential axis ratio $\sim 1-\epsilon_0 \sim 0.9$.  $\epsilon_0$ needs
to be small so as to guarantee
that the volume density of the model is positive everywhere at all time.
We set $T_G/4=4$Gyr, 
a reasonable time scale for the growth of the Galactic disk.

Second we consider a Galactic potential where
the minor axis of the potential slowly flips over a time scale $T_F$, 
\beq
\Phi_F(r,\theta,t) = V_0^2 \left[ \log r + {\epsilon_0 \over 2} \cos 2(\theta-{\pi t \over 
T_F})\right].
\eeq
This potential might mimic the tidal harassment
of Local Group galaxies.  We set $T_F=2$Gyr for the time to flip $180^o$.
This type of model has also been examined by Ostriker \& Binney (1989).

For the last case we analyze
a static Galactic potential plus a time-varying perturbation coming from
a massive satellite on a circular orbit with a period $T_P$.  The potential
is
\beq
\Phi_P(r,\theta,t) = V_0^2 \left[\log r + {\epsilon_0 \over 2} \cos 2\theta\right]-{GM_P \over 
\sqrt{|{\bf R}-{\bf R}_P(t)|^2+b_P^2}} ,
\eeq
where we use a Plummer model for the perturber's potential with a 
scale length $b_P$ about half of the tidal radius,
the total mass of the perturber $M_P$ is
taken to be $5\%$ of the Galactic mass enclosed by its circular orbit 
of period $T_P$ and radius $R_P={ V_0 T_P \over 2 \pi}$.
Generally, the orbital plane of the perturber is unrelated to 
that of the stream, but here they are taken to be 
the same plane with both moving in the same sense.  
The perturber is set on an exactly polar, circular orbit with 
a period $T_P=2$Gyr ($R_P=64$ kpc, $M_P=3\times 10^{10}M_\odot$, $2b_P=14$ kpc)
without considering effects such as dynamical friction
and mass-loss.  These parameters might be appropriate for 
a very massive perturber 
such as the progenitor of the Magellanic Clouds (Zhao 1998a,b),
which could significantly ``harass'' all smaller satellites in the halo.

\begin{figure}[]
\epsfysize=70mm
\leftline{\epsfbox{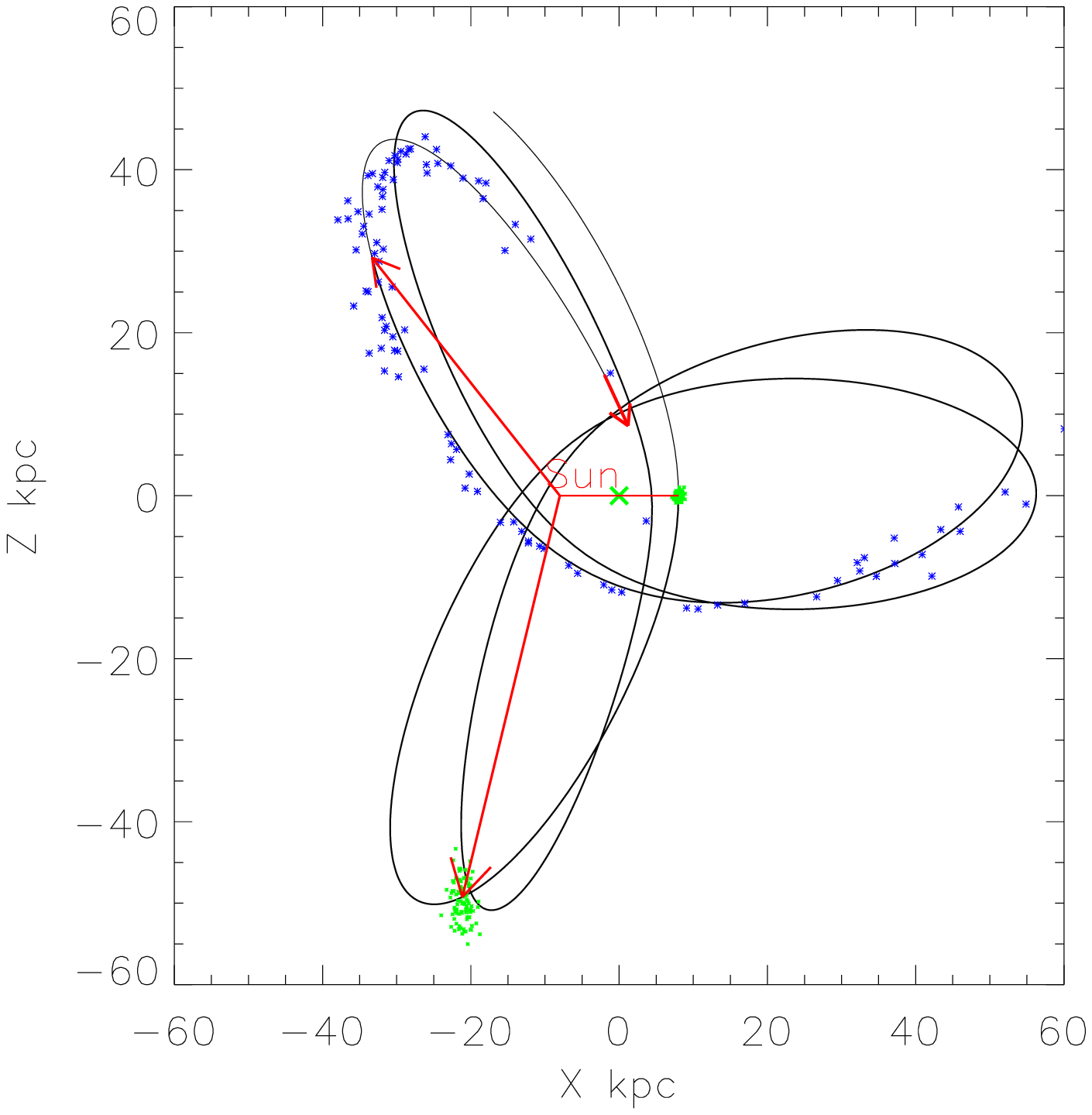}}
\vskip -70mm
\epsfysize=70mm
\rightline{\epsfbox{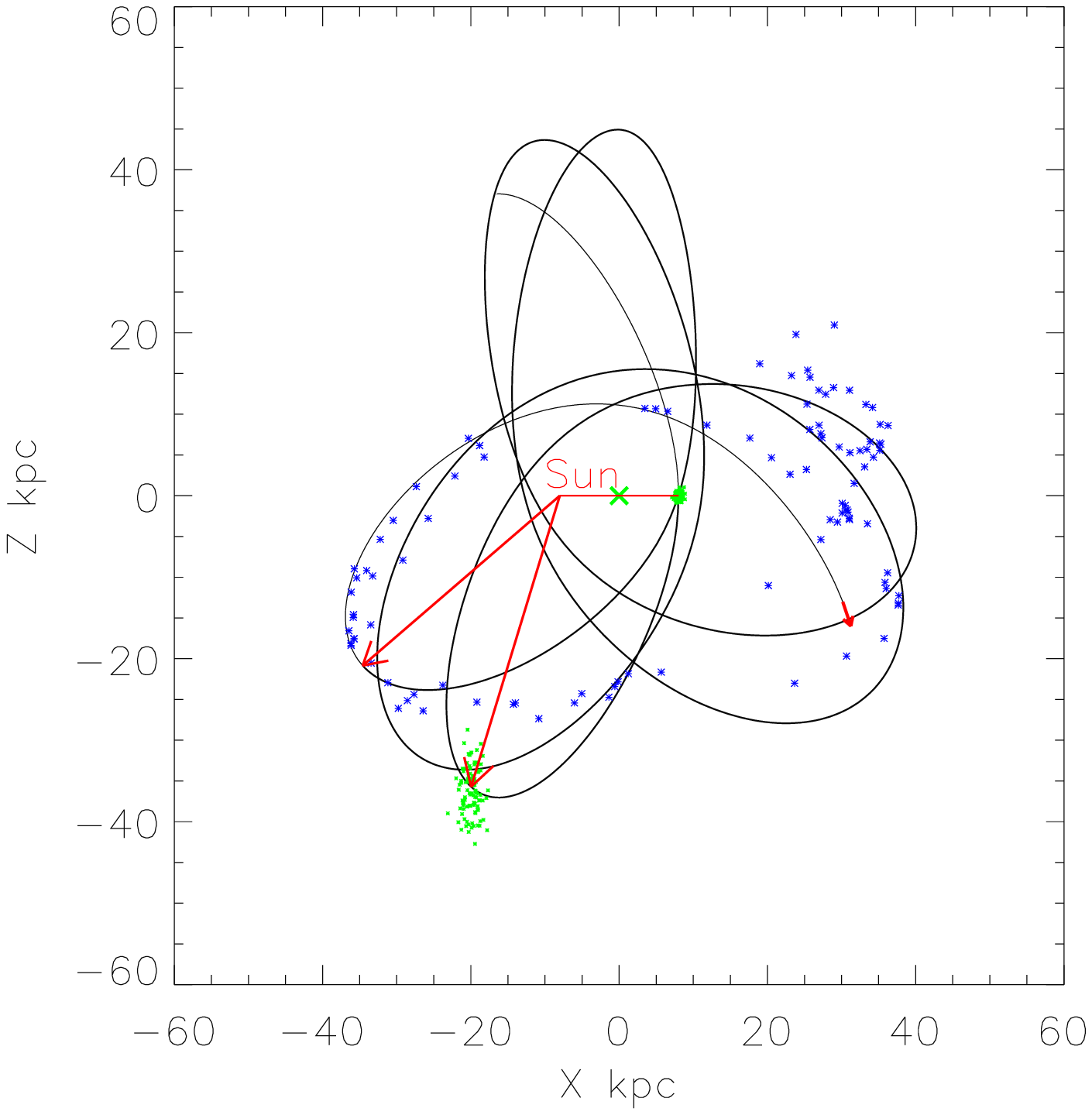}}
\vskip -10mm
\epsfysize=140mm
\leftline{\epsfbox{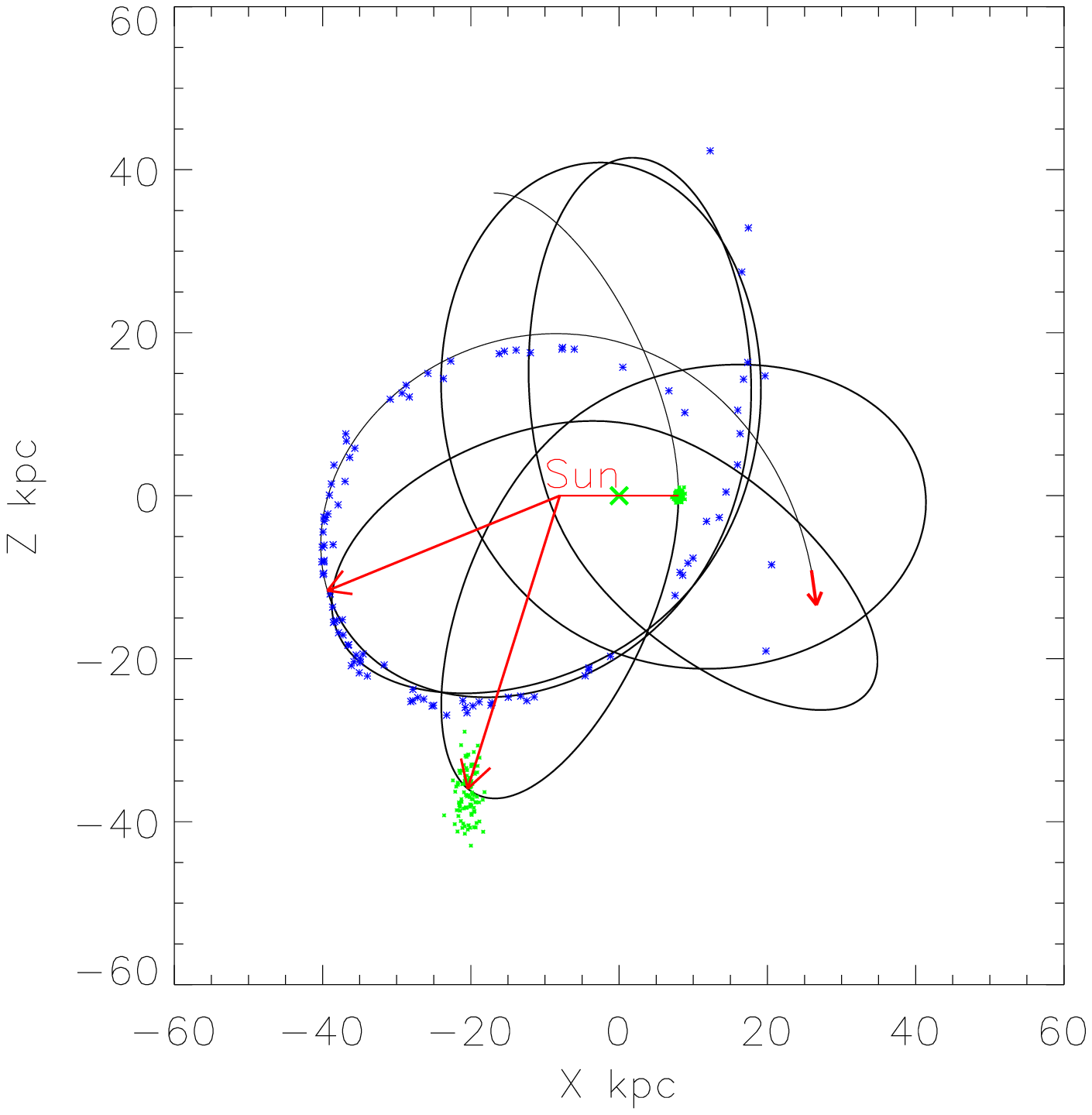}}
\caption{
Simulated orbits of the parent satellite and 
the released stream of about 100 giant stars in potential models 
with a growing disc $\Phi_G$ (upper left panel),
a perturber $\Phi_P$ (upper right) and a flipping disc
$\Phi_F$ (the lower larger panel).  
The Galactic center is marked by a cross.
We assume that the parent satellite is totally
disrupted 4 Gyrs ago at a pericenter $8$ kpc (Sun's mirror image point)
and moving south (down) with a velocity $380\kms$.  The stubby
debris tail shortly ($0.3$ Gyrs) after the disruption is also shown
together with the well-developed tidal tail 4 Gyrs later; 
the center of the satellite circles around like the arms of a clock,
with its position at different epoch is indicated by the long arrows.  
}
\end{figure}
Fig.1 shows the orbit and morphology of the simulated stream in these
three potentials.  The orbit of the disrupted satellite is chosen so
that the released stream stays in the polar $xz$ plane, which passes
through the location of the
Sun and the Galactic center; the $xyz$ coordinate is defined 
such that the Sun is at $x=-8$ kpc and $y=z=0$.

We then simulate observations of mock data of 100 bright HB and GB
stars convolved with GAIA accuracy.  The particles in the disrupted
satellite are initially distributed with an isotropic Gaussian density
and velocity profile (as in Helmi \& White 1999) with dispersions $0.4$
kpc and $4\kms$ respectively.  These particles are released
instantaneously at the pericenter $8$ kpc from the center $4$ Gyrs
ago.  These parameters might be most relevant for satellites such as
the progenitor of the Sagittarius stream.  The parallax and radial
velocity from GAIA will not be very constraining for stream stars
beyond 10 kpc ($100\mu$as in parallax) because of rapid growth of
error bars with magnitude of the stars (Lindegren, private
communications).  Here we adopt a simple parameterization for the
error of the sample stars.  We consider only horizontal branch stars,
for which the errors in parallax ($\pi$ in $\mu$as), proper motion
($\mu$ in $\mu$as\ yr$^{-1}$) and heliocentric radial velocity ($V_h$
in $\kms$) are functions of the parallax.  We find that this simple
formula
\beq
\sigma_\pi=1.6\sigma_\mu=\sigma_{V_h}=f(\pi), ~~~f(x)=5+50 (50/x)^{1.5},
\eeq
approximates the GAIA specifications fairly well.

Nevertheless, Fig. 2 and Fig. 3 show that the distribution of the
proper motions of the stream remains narrow (as shown by the proper
motion vs. proper motion diagram and the position-proper motion
diagram) after taking observational error into account.  The
narrow distribution allows stars in a stream to be selected out from
random field stars, as we argue in the previous section.
\begin{figure}[]
\epsfysize=160mm
\center{\epsfbox{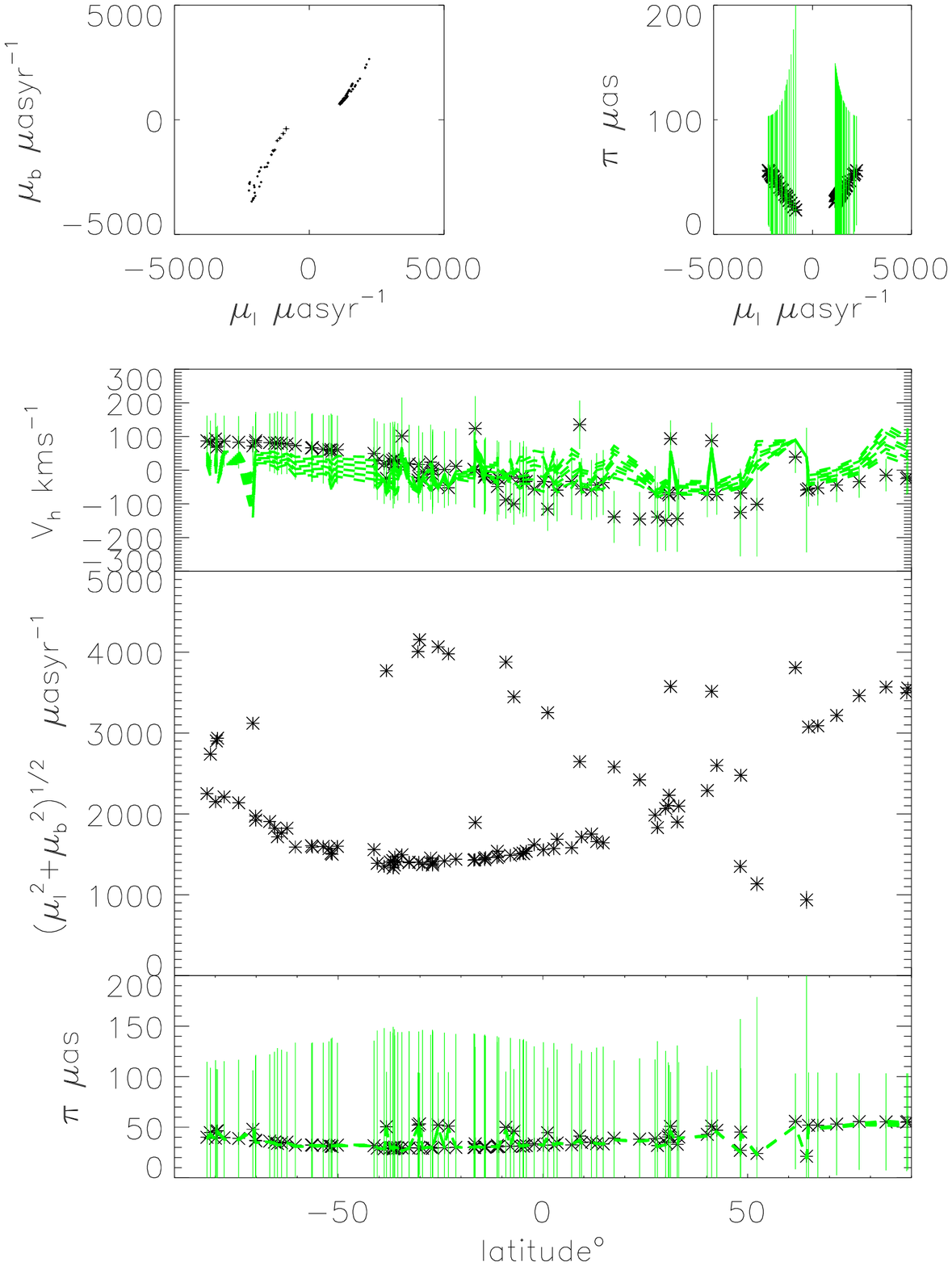}}
\caption{
Mock data of the stream convolved with GAIA accuracy for the potential model
with a flippig disc $\Phi_F$.  
The two small panels at the top are the proper motion
vs. proper motion ($\mu_l$ vs. $\mu_b$) diagram and the (longitude)
proper motion vs. parallax ($\mu_l$ vs $\pi$) diagram; the linear
regression $|\mu_l| \sim 200 \kms/{\rm D\, kpc} \sim 40\pi$ 
manifests the reflex of the Galactic rotation of the Sun 
and the absence of any azimuthal rotation of a polar stream.  
The lower three panels show the heliocentric 
radial velocity, the proper motion, and the parallax across
the stream as functions of the latitude; the GAIA error bars for the
proper motion are typically much smaller than the size of the symbols.  
The two narrow bands bracket the range of parallax and heliocentric velocity
predicted from assuming a constant angular momentum $|{\bf J}|$ and
approximate energy $E$ across the stream; the width of the bands here are
computed from the initial spread of $E$ and ${\bf J}$.
}
\end{figure}

\begin{figure}[]
\epsfysize=160mm
\epsfxsize=60mm
\leftline{\epsfbox{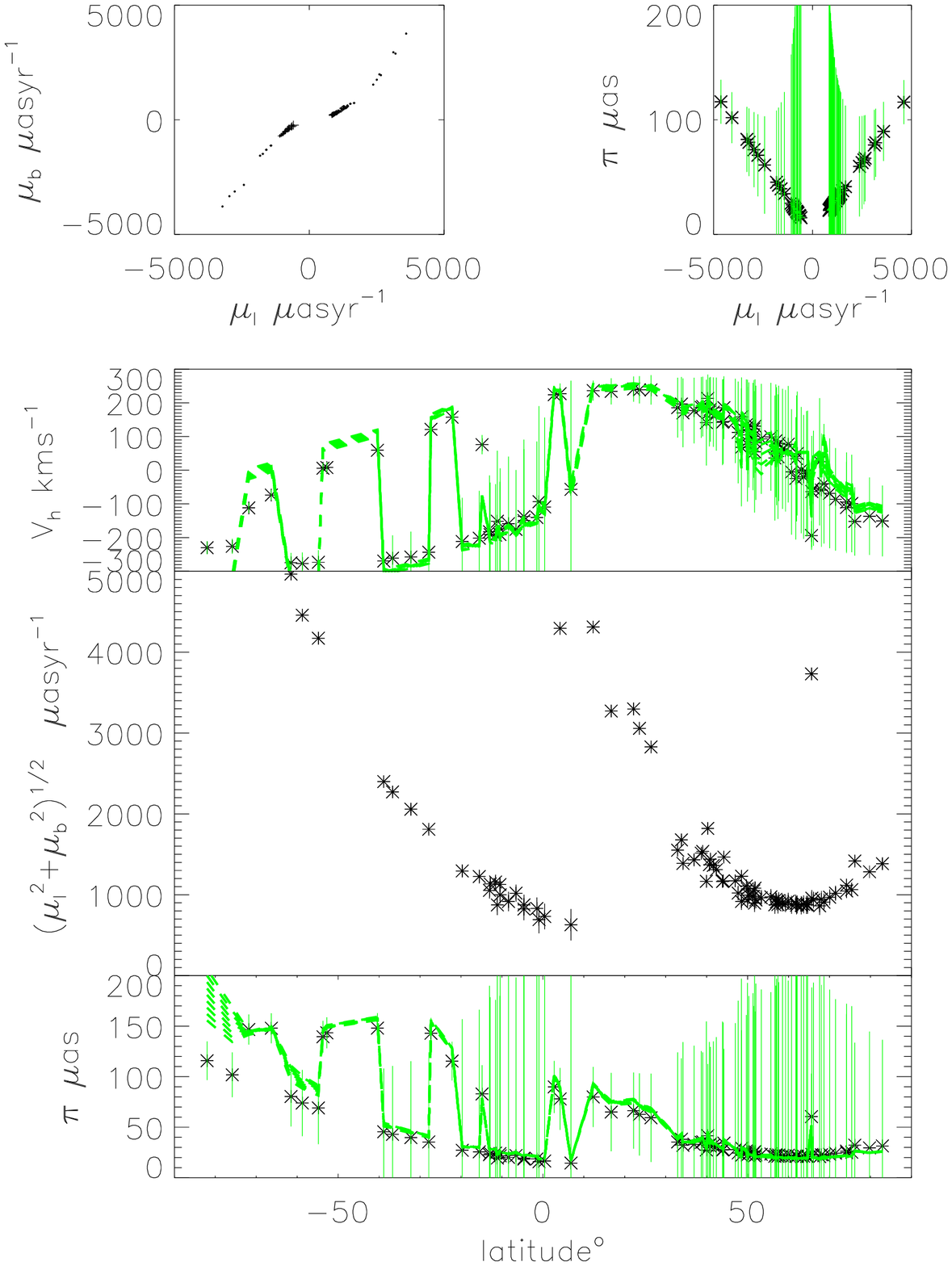}}
\vskip -160mm
\epsfysize=160mm
\epsfxsize=60mm
\rightline{\epsfbox{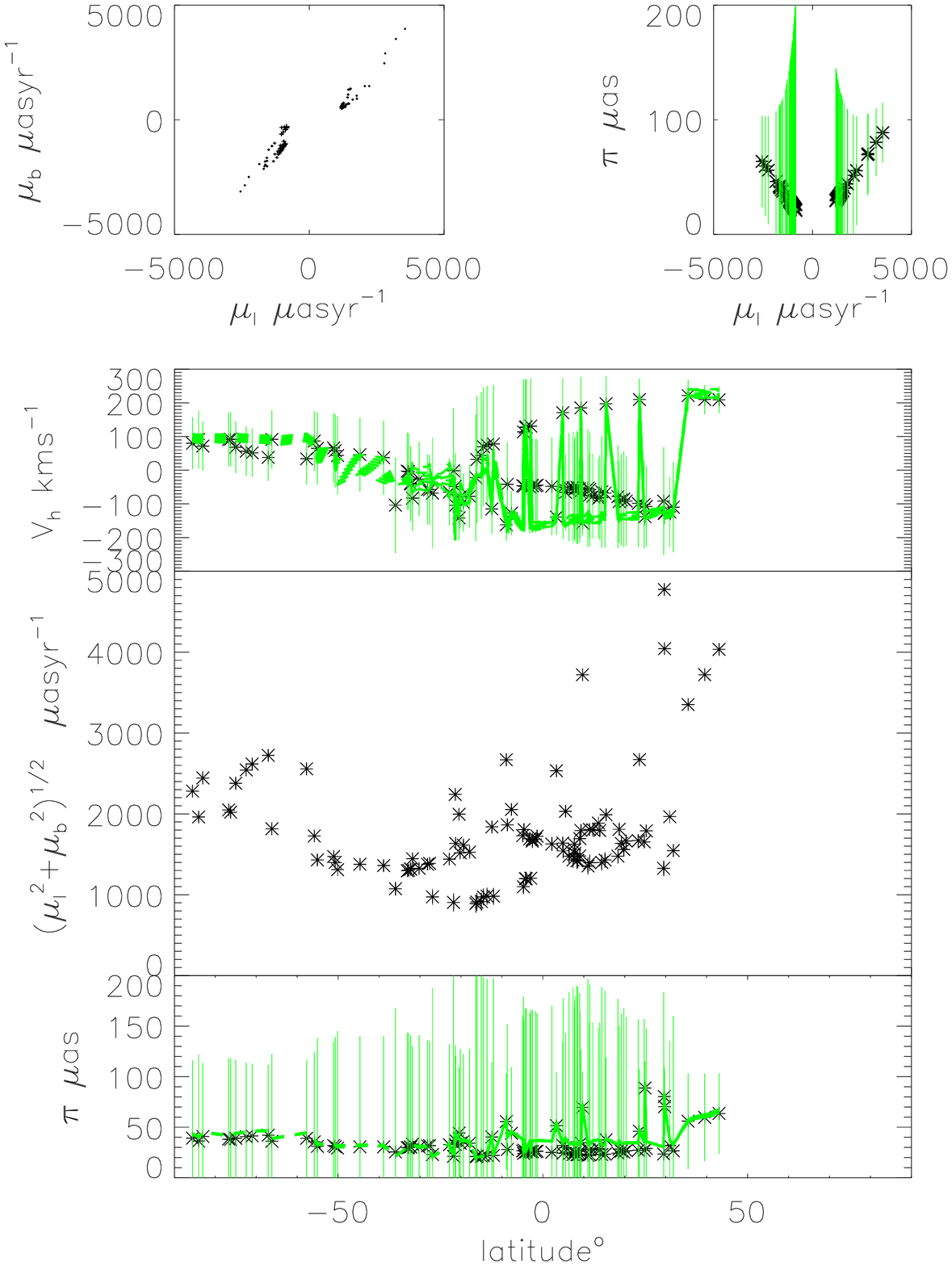}}
\caption{Same as the previous figure, but for models 
with a growing disc $\Phi_G$ (left panels)
and a perturber $\Phi_P$ (right panels).}
\end{figure}

Fig.4 shows the simulated streams in the energy and angular momentum
space for various evolution histories of the Galactic potential.  For
example, in the model where the Galaxy slews, particles move along
straight lines defined by the Jacobian integral $E-{\pi \over T_F}J$,
where $J$ is the angular momentum in the direction around which the
system slews.  By and large the energy $E$ of particles across each
stream is spread out only in a narrow range at each epoch in the three
models; the same holds but to a lesser extent for the angular momentum
vector ${\bf J}$.  This implies that stars in the stream are largely
coeval even in the presence of realistic, moderate evolution of the
Galactic potential.  The energy and angular momentum is also modulated
with particle position in a sinusoidal way across the stream, an
effect which in principle can be used to infer the evolution rate of
the Galactic potential and the flattening of the potential.

\begin{figure}[]
\epsfysize=160mm
\center{\epsfbox{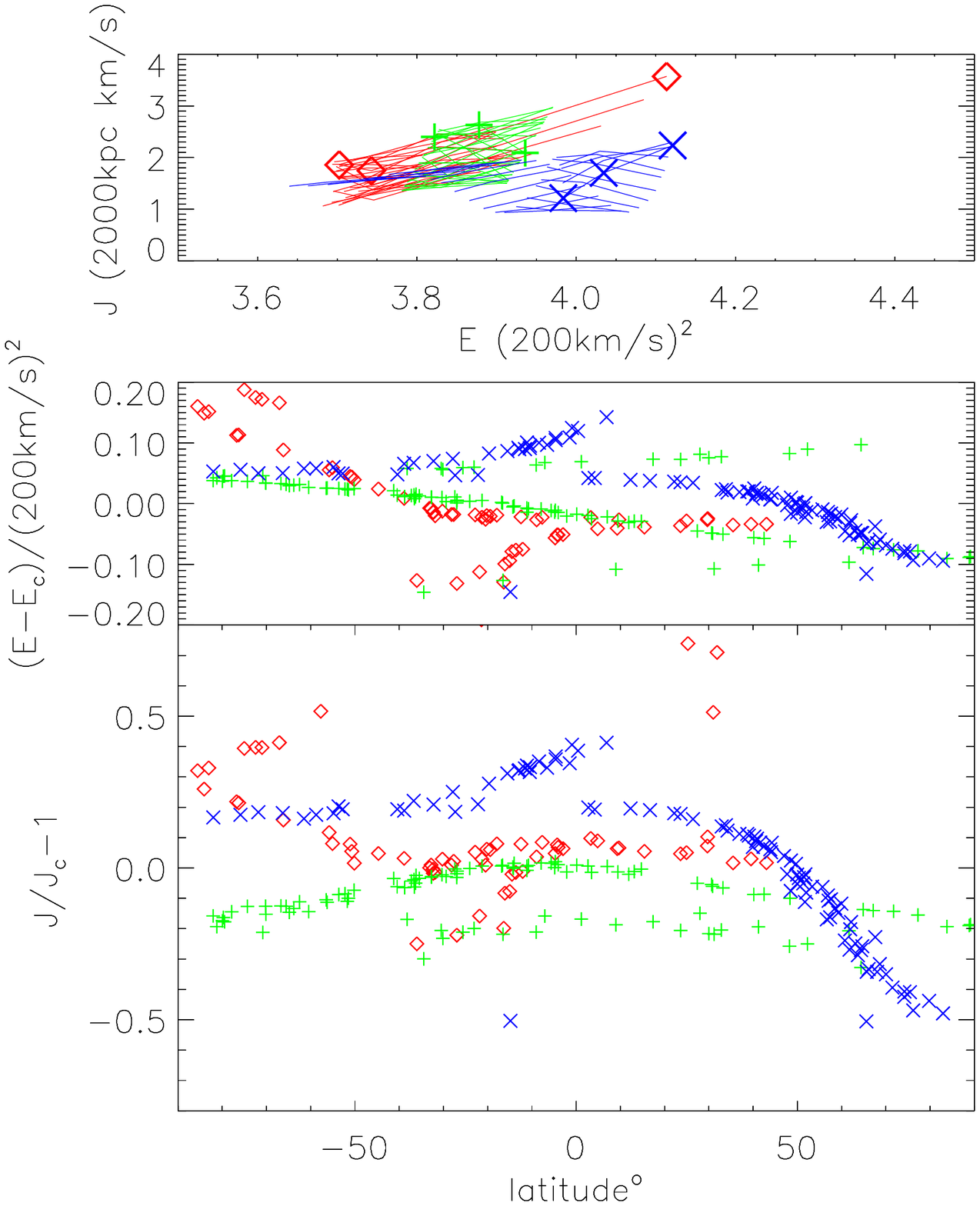}}
\caption{
The lower panels compare the angular momentum and energy along three
streams simulated in three different time-varying potentials.  Each
shaded region in the top panel shows how most ($\sim 70\%$) stars in a
stream evolve in the energy vs. angular momentum plane for one of the
three potentials.  They are shown in steps of about $0.15$ Gyr from 4
Gyrs ago to the present epoch (marked by the symbols).  The plus
symbols are for the flipping Galactic potential with $T_F=2$ Gyrs, the
diamonds are for the case with a massive perturber on a $T_P=2$ Gyrs
orbit, the crosses are for a slowly growing potential with $T_G/4=4$
Gyrs.  $J_c$ and $E_c$ are the present values for the center of mass
of the stream.  }
\end{figure}
One of the challenges of using streams to constrain the potential is
that measurements of both parallax and radial velocity from GAIA are
likely dominated by noise for the fainter ($M_V \sim 20$ mag), more
distant ($\sim 50$ kpc) members of a halo stream (cf. Table 1).
Fortunately, we can use the property of nearly constant angular
momentum and energy across the stream to predict the missing
information with an accuracy often comparable to or better than
directly observable by GAIA.  In essence we apply variations of the
classical method of obtaining ``secular parallaxes''.  The simplest
example is a polar stream, which has no net azimuthal angular
momentum, $J_z \sim 0$, so parallaxes can be recovered (to about
$10\mu$as accuracy) from the solar reflex motion of the stream in the
longitude direction as shown by the linear regression $\pi \sim
|\mu_l|/40$ in the top right panel of Fig. 2.  More generally we can
use the property that the total angular momentum and an approximate
energy are roughly constant, i.e.,
\beq
{\bf J} = {\bf r \times V} \sim {\rm constant},~~~ 
E \sim (200\kms)^2 \log r + {1 \over 2} {\bf V}^2 \sim {\rm constant}
\eeq
to predict both parallax and heliocentric velocity; here we simply
pretend that the Galactic potential is spherical.  Surprisingly, these
very rough approximations often yield fairly accurate parallaxes
($\sim 10\%$) and heliocentric velocities ($\sim 30\kms$) as shown
by the narrow bands in Fig. 2 and Fig. 3; predictions tend to be poorer
for particles in the (anti-)center direction, because it is where the
angular momentum ${\bf J}$ becomes insensitive to the heliocentric velocity.
The predicted velocities and parallaxes are testable with direct observations
at least for the nearby brighter members of a stream.

\begin{figure}[]
\epsfysize=160mm
\center{\epsfbox{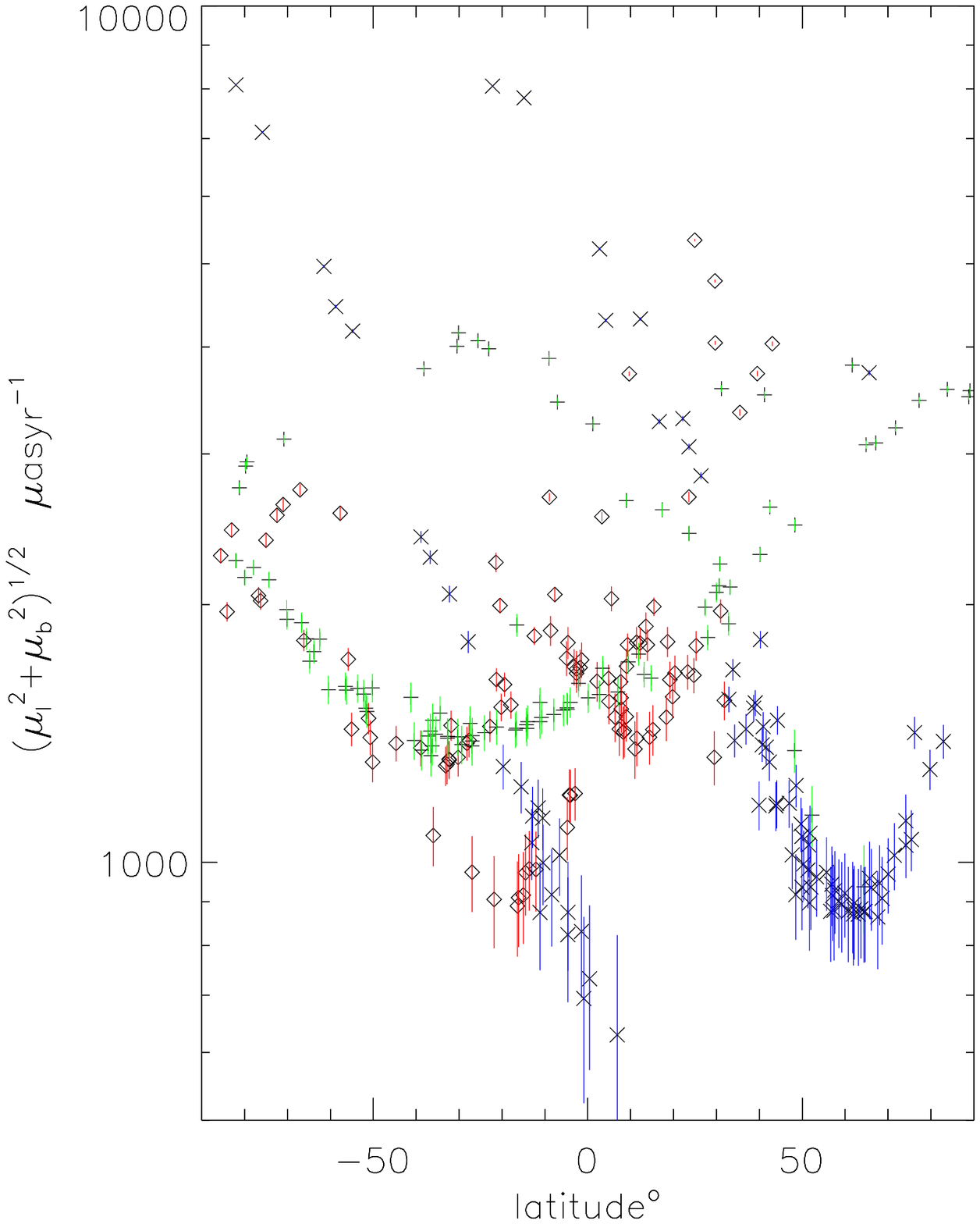}}
\caption{
Mock data for the norm of the proper motion vector across three streams
simulated with GAIA accuracy.  The three streams are generated
in three different evolution histories of the Galactic potential.
Symbols have the same meaning as in the previous figure.
}
\end{figure}
Most relevant to constraining the evolution of the Galaxy is that the
differences in the position-proper motion diagram (Fig. 5), reflected
by these different evolution scenarios, are at a level clearly
resolvable with GAIA accuracy.  We caution that the examples shown
here are {\it biased} towards nicely-structured and extended streams.
These happen quite often for models where the Galaxy simply slews, but
not for models involving a growing disk or a massive perturber as we
increase the amplitude of the perturbation.  We have also run
simulations with various parameters for the satellite orbit and
initial size and the Galactic potential.  The structure of a stream
can become very noisy for highly eccentric orbits with a pericenter
smaller than 8 kpc and/or for potentials where the temporal
fluctuation of the rotation curve is greater than 10\%.  These noisy
structures as a result of strong evolution can be challenging to
detect.  Nevertheless we conclude that tidal streams are excellent
tracers of the Galactic potential as long as a stream maintains a cold
spaghetti-like structure, in particular, the results of Johnston et
al. (1999) and Helmi et al. (1999) for static Galactic potentials are
likely to be largely generalizable to moderately time-evolving
potentials.  However, perhaps the most exciting implication of these
preliminary results is that by mapping the proper motions along the
debris with SIM or GAIA we could eventually set limits on the rate of
evolution of the Galactic potential, and distinguish among scenarios
of Galaxy formation.

\vskip 1cm

HSZ thanks Amina Helmi and Tim de Zeeuw for many helpful comments on
an earlier version.


\clearpage
\begin{thebibliography}{}
\bibitem{} Gilmore G. et al. 1998, Proc SPIE Conference 3350, 
March 1998 (in press) Astronomical Interferometry
\bibitem{} Grillmair C.J. et al. 1998, Workshop on Galactic Halos, Santa Cruz, ed. D. Zaritsky (astro-ph/9711223)
\bibitem{} Helmi A. \& White S.D.M. 1999, MNRAS, submitted
\bibitem{} Helmi A., Zhao H.S., \& de Zeeuw P.T. 1999, this volume
\bibitem{} Ibata R.,  Gilmore G. \& Irwin M.J. 1994, Nature, 370, 194 
\bibitem{} Irwin M. J. \& Hatzidimitriou D. 1995, MNRAS, 277, 1354
\bibitem{} Johnston K.V., Zhao H.S., Spergel D.N., Hernquist L. 1999, 
ApJ Letters, in press.
\bibitem{} Murai T. \& Fujimoto M. 1980, PASJ, 32, 581
\bibitem{} Ostriker E. \& Binney J.J. 1989, MNRAS, 237, 785
\bibitem{} Putman M. et al. 1999, this volume
\bibitem{} Unwin S.C. et al. 1998, http://sim.jpl.nasa.gov/
\bibitem{} Zhao H.S., 1998a, MNRAS, 294, 139
\bibitem{} Zhao H.S., 1998b, ApJ 500, L49
\end{thebibliography}
\end{document}